\documentclass{emulateapj}

\usepackage{graphicx}
\usepackage{amsmath}
\usepackage{natbib}
\usepackage{amssymb}

\usepackage[breaklinks,colorlinks,urlcolor=blue,citecolor=blue,linkcolor=blue]{hyperref}
\usepackage{hyperref}

\def\LIGO{GW170817}
\newcommand{\eos}{{\scriptstyle\rm EoS}}

\begin{document}

\newcommand{\be}{\begin{equation}}
\newcommand{\ee}{\end{equation}}

\newcommand{\BM}[1]{\textcolor{cyan}{#1}}

\title{Constraining the Maximum Mass of Neutron Stars From Multi-Messenger Observations of GW170817}

\author{Ben Margalit \& Brian D.~Metzger}
\altaffiltext{1}{Department of Physics and Columbia Astrophysics Laboratory, Columbia University, New York, NY 10027, USA.  email: btm2134@columbia.edu, bdm2129@columbia.edu}

\begin{abstract}
We combine electromagnetic (EM) and gravitational wave (GW) information on the binary neutron star (NS) merger \LIGO~in order to constrain the radii $R_{\rm ns}$ and maximum mass $M_{\rm max}$ of NSs.  \LIGO~was followed by a range of EM counterparts, including a weak gamma-ray burst (GRB), kilonova (KN) emission from the radioactive decay of the merger ejecta, and X-ray/radio emission consistent with being the synchrotron afterglow of a more powerful off-axis jet.  The type of compact remnant produced in the immediate merger aftermath, and its predicted EM signal, depend sensitively on the high-density NS equation of state (EOS).  For a soft EOS which supports a low $M_{\rm max}$, the merger undergoes a prompt collapse accompanied by a small quantity of shock-heated or disk wind ejecta, inconsistent with the large quantity $\gtrsim 10^{-2}M_{\odot}$ of lanthanide-free ejecta inferred from the KN.  On the other hand, if $M_{\rm max}$ is sufficiently large, then the merger product is a rapidly-rotating supramassive NS (SMNS), which must spin-down before collapsing into a black hole.  A fraction of the enormous rotational energy necessarily released by the SMNS during this process is transferred to the ejecta, either into the GRB jet (energy $E_{\rm GRB}$) or the KN ejecta (energy $E_{\rm ej}$), also inconsistent with observations.  By combining the total binary mass of \LIGO~inferred from the GW signal with conservative upper limits on $E_{\rm GRB}$ and $E_{\rm ej}$ from EM observations, we constrain the likelihood probability of a wide-range of previously-allowed EOS.  These two constraints delineate an allowed region of the $M_{\rm max}-R_{\rm ns}$ parameter space, which once marginalized over NS radius places an upper limit of $M_{\rm max} \lesssim 2.17M_{\odot}$ (90\%), which is tighter or arguably less model-dependent than other current constraints.
\end{abstract}

\maketitle

\section{Introduction}

On August 17, 2017, the Advanced LIGO and Virgo network of gravitational wave (GW) observatories discovered the inspiral and coalescence of a binary neutron star (BNS) system \citep{LIGO+17DISCOVERY}, dubbed \LIGO.  The measured binary chirp mass was $\mathcal{M}_{\rm c} = 1.118^{+0.004}_{-0.002}M_{\odot}$, with larger uncertainties on the mass of the individual neutron star (NS) components and total mass of $M_{1} = 1.36$-$1.60 M_{\odot}$, $M_{2} = 1.17$-$1.36M_{\odot}$, and $M_{\rm tot} = M_{1}+M_{2} = 2.74^{+0.04}_{-0.01}M_{\odot}$, respectively. These masses are derived under the prior of low dimensionless NS spin ($\chi \lesssim 0.05$), 
characteristic of Galactic BNS systems.

The electromagnetic follow-up of \LIGO~was summarized in \citet{LIGO+17CAPSTONE}.  The {\it Fermi} and INTEGRAL satellites discovered a sub-luminous gamma-ray burst (GRB) with a sky position and temporal coincidence within $\lesssim$ 2 seconds of the inferred coalescence time of \LIGO~\citep{Goldstein+17,Savchenko+17,LIGO+17FERMI}.  Eleven hours later, an optical counterpart was discovered \citep{Coulter+17,Allam+17,Yang+17,Arcavi+17b,Tanvir&Levan17,Lipunov+17} with a luminosity, thermal spectrum, and rapid temporal decay consistent with those predicted for ``kilonova" (KN) emission, powered by the radioactive decay of heavy elements synthesized in the merger ejecta \citep{Li&Paczynski98,Metzger+10}.  The presence of both early-time visual (``blue") emission \citep{Metzger+10} which transitioned to near-infrared (``red") emission \citep{Barnes&Kasen13,Tanaka&Hotokezaka13} at late times requires at least two distinct ejecta components consisting, respectively, of light and heavy r-process nuclei  (e.g.~\citealt{Cowperthwaite+17,Nicholl+17,Chornock+17,Kasen+17,Drout+17,Kasliwal+17}).  
Rising X-ray \citep{Troja+17,Margutti+17} and radio \citep{Hallinan+17,Alexander+17} emission was observed roughly two weeks after the merger, consistent with delayed onset of the synchrotron afterglow of a more powerful relativistic GRB whose emission was initially relativistically beamed away from our line of sight (e.g.~\citealt{vanEerten&MacFadyen11}).

\begin{figure*}[!t]
\includegraphics[width=1.0\textwidth]{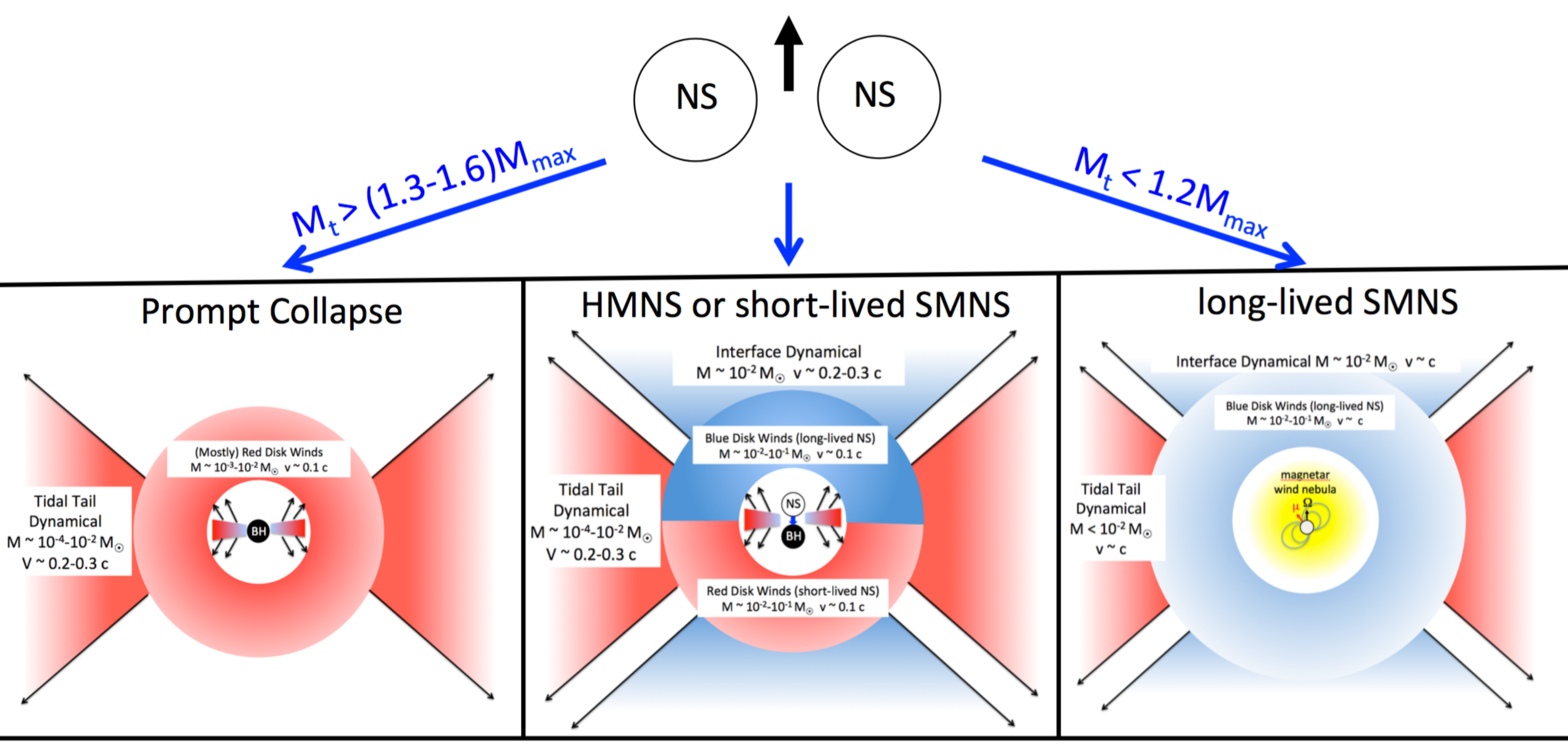}
\caption{\footnotesize The strength of the red and blue KN signatures of a BNS merger depends on the compact remnant which forms immediately after the merger; the latter in turn depends on the total mass of the original binary or its remnant, $M_{\rm tot}$, relative to the maximum NS mass, $M_{\rm max}$.  A massive binary ($M_{\rm tot} \gtrsim 1.3-1.6M_{\max}$) results in a prompt collapse to a BH; in such cases, the polar shock-heated ejecta is negligible and the accretion disk outflows are weakly irradiated by neutrinos, resulting in a primarily red KN powered by the tidal ejecta (left panel).  By contrast, a very low mass binary $M_{\rm tot} \lesssim 1.2M_{\rm max}$ creates a long-lived SMNS, which imparts its large rotational energy $\gtrsim 10^{52}$ erg to the surrounding ejecta, imparting relativistic expansion speeds to the KN ejecta or producing an abnormally powerful GRB jet (right panel).  In the intermediate case, $1.2M_{\rm max} \lesssim M_{\rm tot} \lesssim 1.3-1.6M_{\rm max}$ a HMNS or short-lived SMNS forms, which produces both blue and red KN ejecta expanding at mildly relativistic velocities, consistent with observations of \LIGO. }
\label{fig:schematic}
\end{figure*}

The discovery of \LIGO~implies a BNS rate of $\mathcal{R}_{\rm BNS} = 1540^{+3200}_{-1220}$ Gpc$^{-3}$ yr$^{-1}$, corresponding to $\approx 6-120$ BNS mergers per year once LIGO/Virgo reach design sensitivity \citep{LIGO+17DISCOVERY}.  This relatively high rate bodes well for the prospects of several scientific objectives requiring a large population of GW detections, such as ``standard siren" measurements of the cosmic expansion history \citep{Holz&Hughes05,Nissanke+10,LIGO+17H0} or as probes of the equation of state (EOS) of NSs \citep[e.g.][]{Read+09,Hinderer+10,Bauswein&Janka12}.

Uncertainties in the EOS limit our ability to predict key properties of NSs, such as their radii and maximum stable mass \citep[e.g.][]{Ozel&Freire16}.  Methods to measure NS radii from GWs include searching for tidal effects on the waveform during the final stages of the BNS inspiral \citep{Hinderer+10,Damour&Nagar10,Damour+12,Favata14,Read+13,DelPozzo+13,Agathos+15,Lackey&Wade15, Chatziioannou+15} and for quasi-periodic oscillations of the post-merger remnant \citep[e.g.][]{Bauswein&Janka12,Bauswein+12a,Clark+14,Bauswein&Stergioulas15,Bauswein+16}. 
Searches on timescales of tens of ms to $\lesssim 500 \, {\rm s}$ post-merger revealed no evidence for such quasi-periodic oscillations in the \LIGO \citep{LIGO+17DISCOVERY}. 

While the radii of NS are controlled by the properties of the EOS at approximately twice the nuclear saturation density, the maximum stable mass \citep{Lattimer&Prakash01}, $M_{\rm max}$ instead depends on the very high density EOS \citep[around 8 times the saturation density;][]{Ozel&Psaltis09}.  Observations of two pulsars with gravitational masses of $1.93\pm 0.07M_{\odot}$ \citep{Demorest+10,Ozel&Freire16} or $2.01\pm 0.04M_{\odot}$ \citep{Antoniadis+13} place the best current lower bounds .  However, other than the relatively unconstraining limit set by causality, no firm theoretical or observational upper limits exist on $M_{\rm max}$.  Indirect, assumption-dependent limits on $M_{\rm max}$ exist from observations of short GRBs (e.g.~\citealt{Lasky+14,Lawrence+15,Fryer+15,Piro+17}) and by modeling the mass distribution of NSs (e.g.~\citealt{Antoniadis+16,Alsing+17}).   

\begin{figure}[!t]
\includegraphics[width=0.5\textwidth]{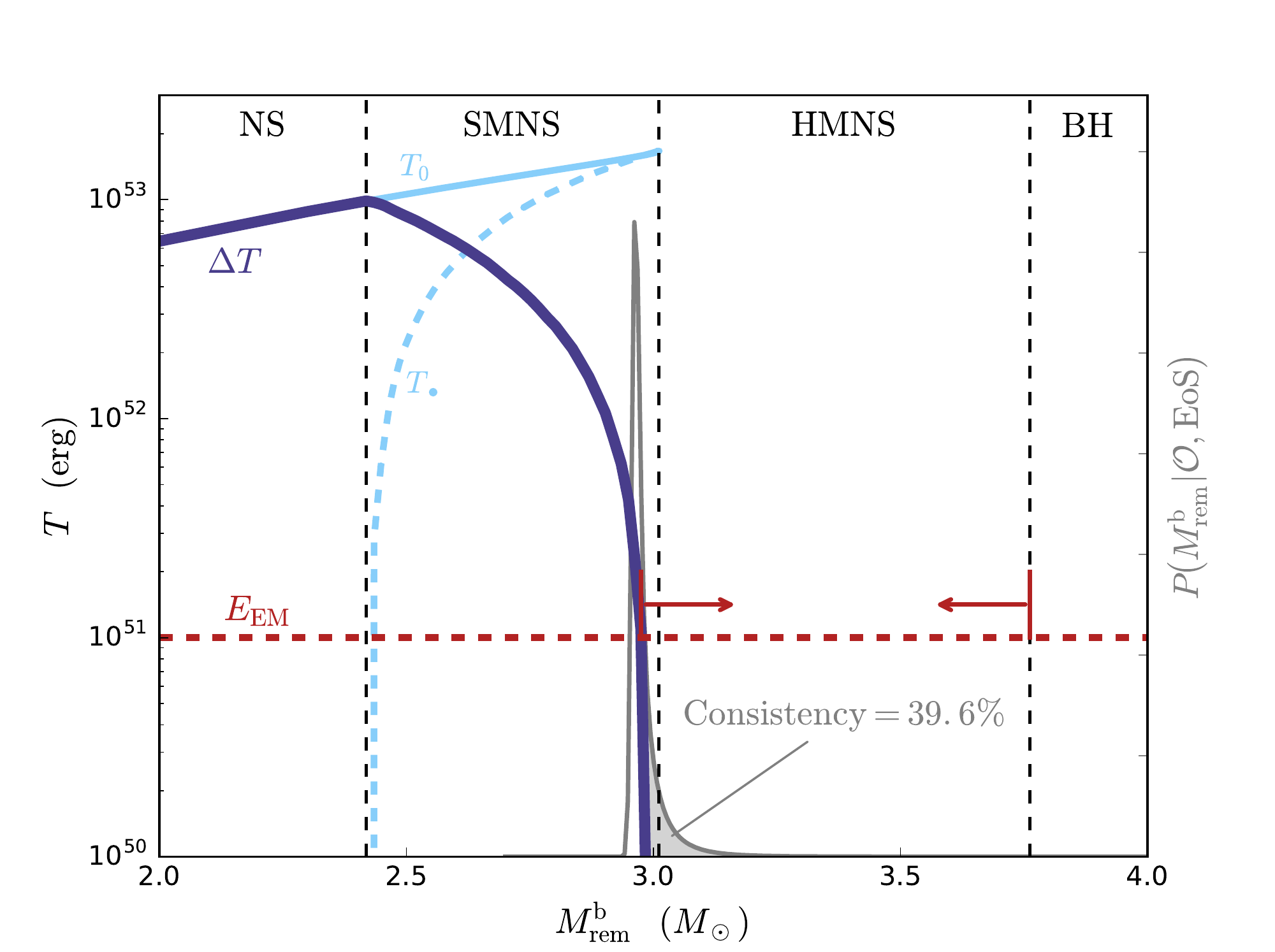}
\hspace{0.0cm}
\caption{
The maximum extractable rotational energy of the merger remnant
$\Delta T = T_{0}-T_{\bullet}$ (Eq.~\ref{eq:DeltaT}) 
is shown as a dark-blue solid curve for a sample EOS.
Vertical dashed curves demarcate the range of baryonic remnant masses $M_{\rm rem}^{\rm b}$ for which the immediate post-merger compact object is a stable NS, SMNS, HMNS, or a BH (prompt-collapse).  A horizontal red dashed curve shows the maximal energy transferred to the environment of the merger consistent with EM observations of \LIGO~for the GRB and KN emission.  
The parameter space where $\Delta T \gg E_{\rm EM}$ is thus ruled-out. The prompt-collapse scenario is also ruled out (see text), such that $M_{\rm rem}^{\rm b}$ is constrained within an `allowed' region shown by red arrows.  The grey curve shows the remnant mass probability distribution function (Eq.~\ref{eq:P_Mb}), and the consistency is the integral over this distribution within the allowed region (Eq.~\ref{eq:consistency}; shaded-gray area).}
\label{fig:example}
\end{figure}

Despite the large uncertainties on $M_{\rm max}$, it remains one of the most important properties affecting the outcome of a BNS merger and its subsequent EM signal (Fig.~\ref{fig:schematic}).  If the total binary mass $M_{\rm tot}$ exceeds a critical threshold of $M_{\rm th} \approx kM_{\rm max}$, then the merger product undergoes ``prompt" dynamical-timescale collapse to a black hole (BH) \citep[e.g.][]{Shibata05,Shibata&Taniguchi06,Baiotti+08,Hotokezaka+11}, where the proportionality factor $k \approx 1.3-1.6$ is greater for smaller values of the NS ``compactness", $C_{\rm max} = (GM_{\rm max}/c^{2}R_{1.6})$, where $R_{1.6}$ is the radius of a 1.6$M_{\odot}$ NS \citep[e.g.][]{Bauswein+13}.   For slightly less massive binaries with $M_{\rm tot} \lesssim M_{\rm th}$, the  merger instead produces a {\it hyper-massive} neutron star (HMNS), which is supported from collapse by differential rotation (and, potentially, by thermal support).  For lower values of $M_{\rm tot} \lesssim 1.2M_{\rm max}$, the merger instead produces a  {\it supramassive} neutron star (SMNS), which remains stable even once its differential rotation is removed, as is expected to occur $\lesssim 10-100$ ms following the merger \citep{Baumgarte+00,Paschalidis+12,Kaplan+14}.  A SMNS can survive for several seconds, or potentially much longer, until its rigid body angular momentum is removed through comparatively slow processes, such as magnetic spin-down.  
Finally, for an extremely low binary mass, $M_{\rm tot} \lesssim M_{\rm max}$,
the BNS merger produces an indefinitely stable NS remnant 
\citep[e.g.][]{Bucciantini+12,Giacomazzo&Perna13}.  Figure~\ref{fig:example} shows the baryonic mass thresholds of these possible BNS merger outcomes (prompt collapse, HMNS, SMNS, stable) for an example EOS as vertical dashed lines.

The different types of merger outcomes are predicted to create qualitatively different electromagnetic (EM) signals (e.g.~\citealt{Bauswein+13,Metzger&Fernandez14}). In this {\it Letter}, we combine EM constraints on the type of remnant that formed in \LIGO~with GW data on the binary mass in order to constrain the radii and maximum mass of NSs.

\section{Constraints from EM Counterparts}
\label{sec:injection}

This section reviews what constraints can be placed from EM observations on the energy imparted by a long-lived NS into the non-relativistic KN ejecta ($\S\ref{sec:KN}$) and into the relativistic ejecta of the GRB jet ($\S\ref{sec:GRB}$).  Then in $\S\ref{sec:implications}$ we describe the implications for the type of remnant formed.

\subsection{Kilonova (Non-Relativistic Ejecta)}
\label{sec:KN}

Two sources of neutron-rich ejecta, capable of synthesizing $r$-process nuclei, accompany a BNS merger (\citealt{Fernandez&Metzger16}).  First, matter is ejected on the dynamical timescale, either by tidal forces (e.g.~\citealt{Ruffert+97,Rosswog+99,Radice+16}) or by shock heating at the interface between the merging NSs (e.g.~\citealt{Oechslin+07,Bauswein+13,Hotokezaka+13}).  The tidal matter emerges in the binary equatorial plane and has a low electron fraction, $Y_e \lesssim 0.1-0.2$.  Matter from the shocked interface expands into the polar direction and possesses a higher $Y_{e} \gtrsim 0.25$ \citep{Wanajo+14,Sekiguchi+16}.

Outflows from the accretion torus around the central compact object provide a second important source of ejecta \citep[e.g.][]{Metzger+08b,Dessart+09,Fernandez&Metzger13,Perego+14,Just+15,Siegel&Metzger17,Siegel&Metzger17b}.  The disk outflows typically possess a broad distribution of $Y_{e} \sim 0.1-0.5$, with an average $Y_{e}$ that increases with the lifetime of the HMNS/SMNS, due to neutrino irradiation of the ejecta by the NS \citep{Metzger&Fernandez14,Perego+14,Martin+15}.

The KN following \LIGO~showed evidence for two distinct emitting ejecta components \citep{Cowperthwaite+17,Kasen+17,Tanaka+17}.  The early $\lesssim 2$ day timescale ``blue" emission phase requires an ejecta mass of $M_{\rm ej}^{\rm blue} \approx 1\times 10^{-2}M_{\odot}$ of lanthanide-free ejecta ($Y_{e} \gtrsim 0.25$) with a mean velocity of $v_{\rm ej}^{\rm blue} \approx 0.2-0.3c$ \citep{Nicholl+17}.  The comparatively ``red" emission seen at later times requires $M_{\rm ej}^{\rm red} \approx 4-5 \times 10^{-2}M_{\odot}$ of lanthanide-rich ejecta ($Y_{e} \lesssim 0.25$) with $v_{\rm ej}^{\rm red} \approx 0.1-0.2 c$ \citep{Chornock+17}.  The total kinetic energy of the ejecta is therefore approximately $E_{\rm KN} \approx M_{\rm ej}^{\rm blue}(v_{\rm ej}^{\rm blue})^{2}/2 + M_{\rm ej}^{\rm red}(v_{\rm ej}^{\rm red})^{2}/2\approx  1.0\times 10^{51}$ erg.  

\subsection{Gamma-Ray Burst (Relativistic Ejecta)}
\label{sec:GRB}

The radiated gamma-ray energy from \LIGO, and the kinetic energy of its afterglow if it originates from an on-axis GRB jet, were several orders of magnitude lower than those of cosmological short GRBs \citep{Goldstein+17,LIGO+17FERMI,Fong+17}.  This could indicate that we are observing the GRB jet well outside of its core \citep[e.g.][]{Kathirgamaraju+17,Lazzati+17}.  The delayed rise of synchrotron X-ray and radio emission is consistent with the afterglow from a much more powerful relativistic jet pointed away from our line of sight \citep{Troja+17,Hallinan+17,Margutti+17,Alexander+17,Evans+17,Haggard+17}.  However, for observing viewing angles relative to the binary axis inferred from the GW data and host galaxy, $\theta_{\rm obs} \approx 11-33^{\circ}$ \citep{LIGO+17H0}, the inferred kinetic energy of an off-axis GRB jet is $E_{\rm GRB} \lesssim 10^{50}$ erg (e.g~\citealt{Alexander+17,Margutti+17}), within the range of inferred properties of normal cosmological SGRB jets \citep{Berger14}.  

The production of a GRB may indicate that a BH formed (e.g.~\citealt{Lawrence+15,Murguia-Berthier+14}), in which case the GRB's delay of $\lesssim 2$ s following the merger implicates a remnant that either underwent prompt collapse to a BH, or formed a short-lived HMNS or SMNS.   Late-time X-ray emission observed after many short GRBs has been suggested to indicate the presence of a long-lived magnetar \citep[e.g.][]{Metzger+08,Rowlinson+13}, raising doubt about whether BH formation is a strict requirement to produce a GRB.  However, \LIGO~showed no evidence for temporally extended high-energy emission \citep{LIGO+17FERMI}.

\subsection{Constraints on the Merger Remnant in \LIGO}
\label{sec:implications}

The KN emission from \LIGO~tightly constrains the type of compact remnant that formed in the merger event (Fig.~\ref{fig:schematic}).  Prompt collapse to a BH ($M_{\rm tot} \gtrsim M_{\rm th}$) is disfavored by the quantity of the blue KN ejecta.  General relativistic numerical simulations show that mergers with prompt collapses eject only a small quantity $\lesssim 10^{-4}-10^{-3}M_{\odot}$ of matter from the merger interface \citep[e.g.][]{Hotokezaka+11}, inconsistent with the inferred value $M_{\rm ej}^{\rm blue} \gtrsim 10^{-2}M_{\odot}$ for \LIGO.  The accretion disk outflows can also contribute; however, the wind ejecta with $Y_{e} \gtrsim 0.25$ is only a fraction of the initial torus, which is already small $\lesssim 0.01-0.02M_{\odot}$ for prompt collapse \citep{Ruffert&Janka99,Shibata&Taniguchi06,Oechslin&Janka07}.  
Furthermore, the predicted velocities of the disk winds $\sim 0.03-0.1c$ (e.g.~\citealt{Fernandez&Metzger13,Just+15}) are lower than the velocities $\gtrsim 0.2-0.3c$ inferred for the blue KN of \LIGO~(e.g.~\citealt{Nicholl+17}).

A HMNS remnant, due to its longer lifetime $\gtrsim 10$ ms, produces a greater quantity of dynamical and disk wind ejecta.  The expansion rate $\sim 0.2-0.3c$ and ejecta mass of $M_{\rm ej}^{\rm blue} \approx 0.01-0.02M_{\odot}$ of the blue KN ejecta inferred for \LIGO~are consistent with the properties of the high-$Y_{e}$ shock-heated dynamical ejecta found by BNS merger simulations (e.g.~\citealt{Sekiguchi+16}), provided that the radius of the NS is relatively small, $R_{\rm ns} \lesssim 11$ km \citep{Nicholl+17}.  The higher quantity and lower velocity of the red KN emission are also broadly consistent with those expected from the outflows of a relatively massive accretion torus $\approx 0.1-0.2M_{\odot}$ (e.g.~\citealt{Siegel&Metzger17}) following the collapse of a relatively short-lived HMNS.

At the other extreme, a long-lived SMNS or indefinitely stable NS remnant is strongly disfavored by the moderate kinetic energy of the observed KN and GRB afterglow.  Even once its differential rotation has been removed, a SMNS possesses an enormous rotational energy, $T \approx 10^{53}$ erg, which is available to be deposited into the post-merger environment.  Not all of this energy is ``extractable" insofar as, even just prior to spinning down to the threshold for collapse, the NS remnant is still rotating quite rapidly.  \citet{Margalit+15} show that the collapse of a SMNS is unlikely to produce a centrifugally-supported accretion disk outside of the innermost stable circular orbit, in which case all of the mass and angular momentum of the star are trapped in the BH.

The extractable rotational energy of a BNS merger remnant is more precisely defined as
\be \label{eq:DeltaT}
\Delta T = T_{\rm 0} - T_{\bullet},
\ee
where $T_{0}$ is the energy available immediately after differential rotation has been removed, and $T_{\bullet}$ is the rotational energy at the point of gravitational collapse to a BH.  We take $T_{0}$ equal to the rotational energy at the mass-shedding limit, a condition which approximates the state of the remnant immediately after differential rotation is removed.  However, the constraints obtained hereafter would be  similar if we had instead taken $T_{0}$ to equal the threshold value $T/|W| \approx 0.14$ for the growth of secular instabilities (e.g.~\citealt{Lai&Shapiro95}).  Fig.~\ref{fig:example} shows that $\Delta T$ rises sharply from zero at the HMNS-SMNS boundary to $\Delta T = T_{\rm 0} \approx 10^{53}$ erg for stable remnants.

The most likely mechanism by which $\Delta T$ is removed, enabling the SMNS to collapse, is the extraction of angular momentum via a magnetized outflow or jet.  MHD BNS merger simulations find that ultra-strong magnetic fields $\gtrsim 10^{15}-10^{16}$ G are generated in the merger remnant (e.g.~\citealt{Kiuchi+14}).  A NS of radius $R_{\rm ns}$, rotation frequency $\Omega = 2\pi/P$, and spin period $P$ loses energy to a magnetic wind at a rate \citep{Spitkovsky06}
\be
\dot{E}_{\rm mag} = \frac{\mu^{2}\Omega^{4}}{c^{3}}(1 + \sin^{2}\chi),
\label{eq:edotmag}
\ee
where $B_{\rm d}$, $\mu = B_{\rm d}R_{\rm ns}^{3}$, and $\chi$ are the surface magnetic dipole field strength, dipole moment, and angle between the rotation and dipole axes, respectively.  Taking $R_{\rm ns} = 12$ km and $\chi = 0$, the SMNS's available rotational energy is removed by magnetic torques on a timescale
\begin{align}
\tau_{\rm sd} = \frac{\Delta T}{\dot{E}_{\rm mag}} 
\approx 24\,{\rm s} \left(\frac{\Delta T}{10^{52}\,{\rm erg}}\right)\left(\frac{B_{\rm d}}{10^{15}\,{\rm G}}\right)^{-2}\left(\frac{P}{0.8\, \rm ms}\right)^{4}.
\end{align}
If we demand that BH formation occur on a timescale of $\tau_{\rm sd} \lesssim 2$ s following the merger in order to explain the observed gamma-ray emission ($\S\ref{sec:GRB}$), then this requires a SMNS remnant with $B_{\rm d} \gg 10^{15}$ G or $\Delta T\ll 10^{53}$ erg.  

\begin{figure}[!t]
\includegraphics[width=0.5\textwidth]{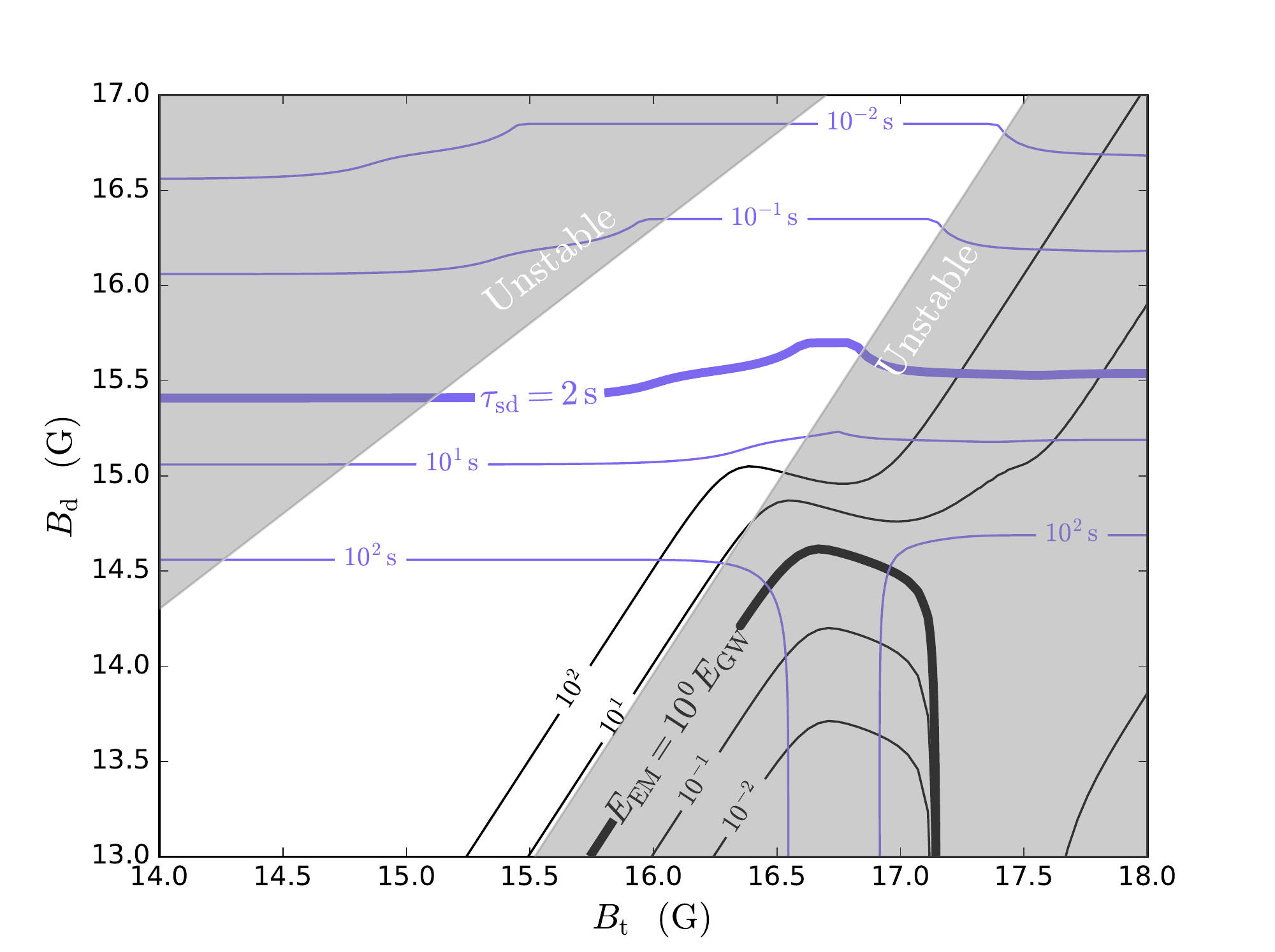}
\caption{\footnotesize Parameter-space of external dipole magnetic field $B_{\rm d}$, responsible for EM spin-down (Eq.~\ref{eq:edotmag}), and internal toroidal field $B_{\rm t}$, which can deform the NS causing GW-driven spin-down. Contours show the spin-down timescale (blue) and ratio of EM to GW extracted spin-down energy (black) calculated by integrating equations for the spin frequency $\Omega$ and misalignment angle $\chi$ as a function of time \citep{Cutler&Jones01,DallOsso+09}. 
The region where GWs could dominate over EM emission falls below the thick black curve, but this region is:
(a) susceptible to magnetic instabilities \citep[grey shaded areas][]{Braithwaite09,Akgun+13}, 
(b) implies long spin-down timescales $\gtrsim 100 \,{\rm s}$ at odds with the detection of a GRB only $2 \, {\rm s}$ after the merger, and 
(c) would produce a strong GW signal.
}
\label{fig:Bfields}
\end{figure}

A SMNS can in principle also spin down through gravitational wave emission, as may result from the quadrupolar moment of inertia induced by a strong {\it interior} magnetic field which is misaligned with the rotation axis \citep[e.g.][]{Stella+05,DallOsso+09,DallOsso+15}.  Figure \ref{fig:Bfields} shows that GW spin-down dominates over magnetic spin-down (Eq.~\ref{eq:edotmag}) only if the interior toroidal magnetic field exceeds the external poloidal one by a factor of $\gtrsim 100$.  However, such a strong toroidal to poloidal field configuration would be unstable (\citealt{Braithwaite09,Akgun+13}; grey shaded region in Fig.~\ref{fig:Bfields}) and would furthermore imply a relatively long collapse time of $\tau_{\rm sd} \gtrsim 100$ s, potentially incompatible with the gamma-ray burst emission observed on a timescale $\lesssim 2$ s ($\S\ref{sec:GRB}$).  It would also produce quasi-periodic GW emission which is not observed in the \LIGO~post-merger signal (albeit with only weakly constraining upper limits; \citealt{LVC17PostMerger}).

In summary, all signs point to \LIGO~having produced a HMNS or very short-lived SMNS remnant.  If the merger had instead produced a long-lived SMNS, then a large fraction of its available rotational energy $\gtrsim 10^{52}$ erg should have been deposited into the merger environment, either into a collimated relativistic jet or shared more equitably with the merger ejecta, on a timescale $\sim \tau_{\rm sd}$.  Such a large energy input is incompatible with the GRB and KN observations of \LIGO.

\section{Constraints on NS Properties}
\label{sec:EOS}

The masses of the binary components inferred for \LIGO, combined with evidence from the KN disfavoring a prompt collapse, places a lower limit on the maximum mass $M_{\rm max}$ of a (slowly rotating) NS.  Likewise, upper limits on the rotational energy injected by a long-lived SMNS place an upper limit on $M_{\rm max}$.  

\begin{deluxetable}{ccccccc}
\tablecaption{EOS Properties and Consistency with EM Observations \label{table:1}}

\tablehead{
\colhead{} & \colhead{$M^{\rm g}_{\rm max}$} & \colhead{$R_{1.3}$} & \colhead{$M^{\rm g}_{\rm smns}$} & \colhead{$\Delta T_{\max}$} & \colhead{Consistency}
\\
\colhead{EOS} & \colhead{$(M_\odot)$} & \colhead{(km)} & \colhead{$(M_\odot)$} & \colhead{($10^{53}$erg)} & \colhead{(\%)}
}
\startdata
MS1 & $2.77$ & $14.9$ & $3.31$ & $1.8$ & $0.0$ \\
MPA1 & $2.45$ & $12.4$ & $2.97$ & $1.8$ & $0.0$ \\
APR3 & $2.37$ & $12.0$ & $2.84$ & $1.7$ & $0.2$ \\
ENG & $2.24$ & $12.0$ & $2.67$ & $1.4$ & $5.2$ \\
WFF2 & $2.20$ & $11.1$ & $2.63$ & $1.6$ & $10.2$ \\
APR4 & $2.19$ & $11.3$ & $2.61$ & $1.5$ & $18.4$ \\
SLy & $2.05$ & $11.8$ & $2.43$ & $1.2$ & $100.0$ \\
H4 & $2.02$ & $14.0$ & $2.38$ & $0.8$ & $100.0$ \\
ALF2 & $1.98$ & $12.7$ & $2.41$ & $0.9$ & $100.0$ \\


GNH3\footnote{\label{fn:tablenote}ruled-out by $2.01 \pm 0.04 M_\odot$ mass of PSR J0348+0432 \citep{Antoniadis+13}} & $1.96$ & $14.3$ & $2.29$ & $0.7$ & $100.0$ \\
ALF4\textsuperscript{\ref{fn:tablenote}} & $1.93$ & $11.5$ & $2.35$ & $1.0$ & $99.8$ \\
BBB2\textsuperscript{\ref{fn:tablenote}} & $1.92$ & $11.2$ & $2.27$ & $1.1$ & $99.4$ \\
MS2\textsuperscript{\ref{fn:tablenote}} & $1.80$ & $14.3$ & $2.10$ & $0.6$ & $99.9$ \\

\enddata

\tablecomments{all EOS are approximated as piecewise broken polytropes \citep{Read+09}}
\end{deluxetable}

\begin{figure}[!t]
\includegraphics[width=0.5\textwidth]{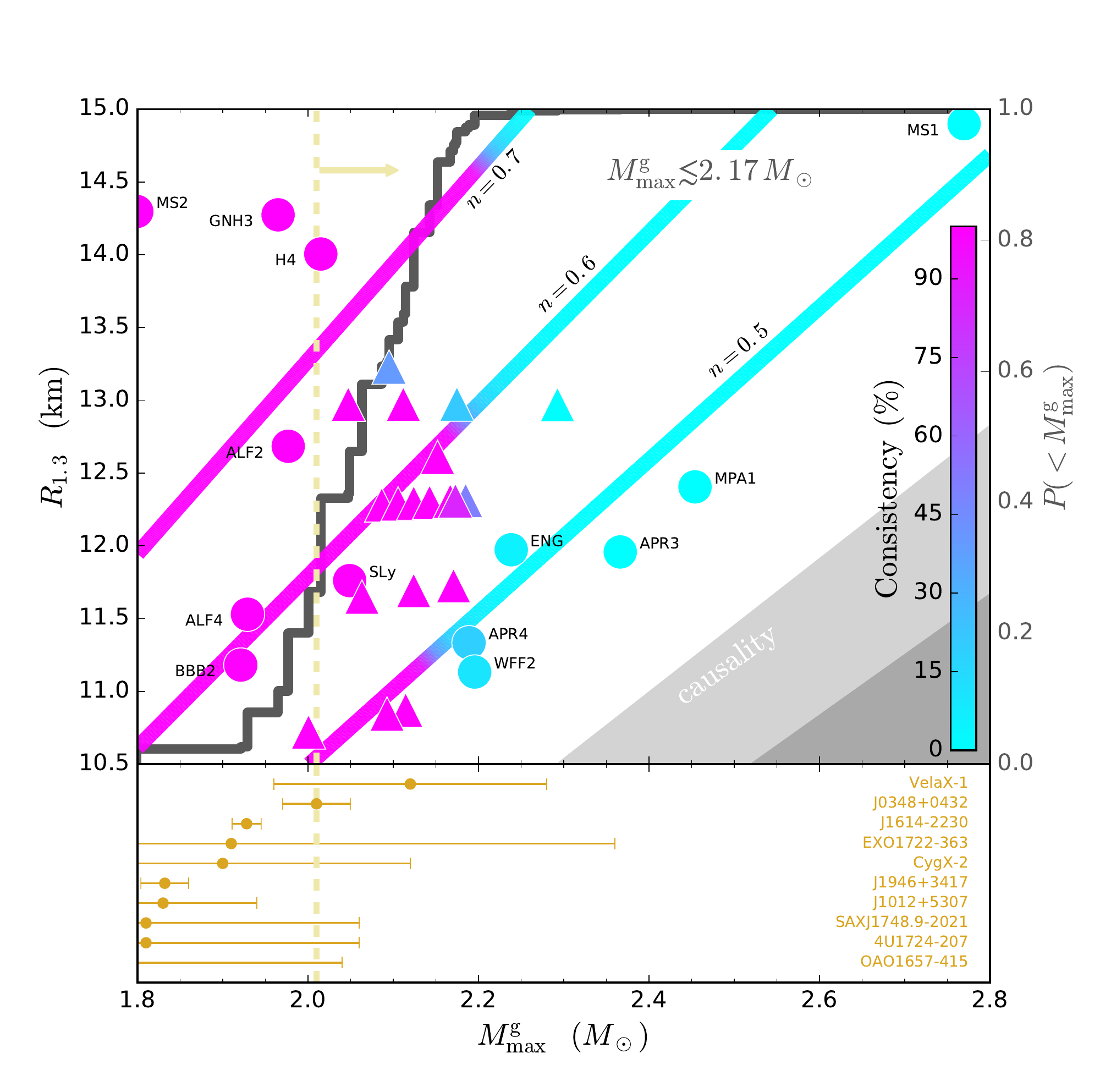}
\caption{\footnotesize Constraints on properties of the NS EOS --- radius of a $1.3 M_\odot$ NS, $R_{1.3}$, and maximal non-rotating gravitational mass, $M^{\rm g}_{\rm max}$ --- based on joint GW-EM observations of \LIGO.  Different EOS are represented as points, the color of which corresponds to the consistency of the given EOS with observational constraints.  
The similarly colored diagonal curves represent polytropic EOSs of index $n$, while the grey shaded regions to the bottom right are ruled out by the requirement of causality (see text).
Clearly, a low NS maximal mass is preferred due to constraints ruling out SMNS formation.
The background grey curve shows the cumulative probability distribution function that the maximum mass $M^{\rm g}_{\rm max}$ is less than a given value (see text),
from which we find $M^{\rm g}_{\rm max} \lesssim 2.17M_{\odot}$ at 90\% confidence.
The bottom panel shows masses of observed Galactic NSs, from which 
a lower limit on $M^{\rm g}_{\rm max}$ 
can be placed (vertical dashed line). }
\label{fig:eos_survey}
\end{figure}

In order to translate GW+EM inferences into constraints on the properties of NSs, we use the {\tt RNS} code \citep{Stergioulas+95} to construct general relativistic rotating hydrostationary NS models for a range of nuclear EOS.
We use the piecewise polytropic approximations to EOS available in the literature provided in \cite{Read+09}  (circles in Fig.~\ref{fig:eos_survey}; summarized in Table~\ref{table:1}), supplemented by EOS constructed from the two-parameter piecewise polytropic parameterization of \cite{Margalit+15} (triangles in Fig.~\ref{fig:eos_survey}).
Our simplified parameterization 
is limited in its ability to model micro-physically motivated EOS with high accuracy, yet allows us to efficiently survey the EOS parameter space,
and we leave it to future work to extend this first analysis with more flexible EOS parameterizations (e.g. \citealt{Raithel+16}).
For each EOS, the uncertainty range of the \LIGO-measured binary gravitational mass \citep{LIGO+17DISCOVERY} translates into a corresponding uncertainty range of baryonic mass, defined by the probability distribution 
\begin{align} \label{eq:P_Mb}
P &\left( M^{\rm b}_{\rm rem} \vert \mathcal{O}, \eos \right) 
= 
\\ \nonumber 
&\int dM^{\rm b}_1 \int dM^{\rm b}_2 \,\,\, 
\delta ( M^{\rm b}_1 + M^{\rm b}_2 - M_{\rm ej} - M^{\rm b}_{\rm rem} )
\\ \nonumber
\times  &P \left( g_\eos(M^{\rm b}_1), g_\eos(M^{\rm b}_2) \vert \mathcal{O} \right)
\left\vert g_\eos^\backprime(M^{\rm b}_1) \right\vert \left\vert g_\eos^\backprime(M^{\rm b}_2) \right\vert .
\end{align}
Here $P \left( M^{\rm g}_1, M^{\rm g}_2 \vert \mathcal{O} \right)$ is the posterior joint probability distribution function of NS gravitational masses inferred from the BNS waveform $\mathcal{O}$ \citep{LIGO+17DISCOVERY}, $M_{\rm ej} = 2 \times 10^{-2} M_\odot$ is a conservative lower limit for the mass loss from the system as inferred from the KN ejecta, and the EOS enters in converting between gravitational and baryonic masses, $M^{\rm g} = g_\eos(M^{\rm b})$.
We approximate the posterior by changing variables to the chirp mass $\mathcal{M}_{\rm c}$ and mass-ratio $q = M^{\rm g}_1 / M^{\rm g}_2$, 
\begin{equation}
P \left( M^{\rm g}_1, M^{\rm g}_2 \vert \mathcal{O} \right) 
= P \left( q, \mathcal{M}_{\rm c} \right) \mathcal{M}_{\rm c}^{-1} q^{6/5} ( 1+q )^{-2/5} ,
\end{equation}
assuming independent asymmetric Gaussian distributions for both $P (\mathcal{M}_{\rm c})$ and $P(q)$, consistent with the median and 90\% quoted confidence levels on $\mathcal{M}_{\rm c}$, $M^{\rm g}_1$, $M^{\rm g}_2$, and $M^{\rm g}_{\rm tot}$.
Specifically, we assume $P \left( q, \mathcal{M}_{\rm c} \right) \propto \exp \left[ -(\mathcal{M}_{\rm c}-\mu_{\mathcal{M}, \pm})^2 / 2 \sigma_\mathcal{M}^2 -(q-\mu_q)^2 / 2 \sigma_q^2 \right]$ for $q \leq 1$ and $P = 0$ otherwise,
with $\mu_q=1, \sigma_q \simeq 0.164, \mu_\mathcal{M} \simeq 1.188 M_\odot$ 
and where $\sigma_{\mathcal{M},\pm} \simeq 2.63 \times 10^{-3} M_\odot$ ($2.07 \times 10^{-3} M_\odot$) for $\mathcal{M}_{\rm c} \geq \mu_\mathcal{M}$ ($\mathcal{M}_{\rm c} < \mu_\mathcal{M}$), respectively.

For each EOS, we then compare the inferred remnant mass to the ``allowed" range between the maximum mass to avoid prompt collapse
(using the relation $M_{\rm th}(R_{1.6},M_{\rm max})$ of \citealt{Bauswein+13}), 
to the minimum baryonic mass which results in a SMNS with an extractable energy $\Delta T$ (Eq.~\ref{eq:DeltaT}) less than the upper limits on the kinetic energy of the KN and GRB emission $E_{\rm EM} = E_{\rm KN} + E_{\rm GRB} \lesssim 10^{51} \, {\rm erg}$.
Integrating the probability distribution of the remnant mass within this allowed range yields the ``consistency''
of the given EOS with the \LIGO~observations,
\begin{equation} \label{eq:consistency}
{\rm Consistency} = \int_{S} P\left( M^{\rm b}_{\rm rem} \vert \mathcal{O}, \eos\right) \, dM^{\rm b}_{\rm rem} ,
\end{equation}
where $S$ is the domain in which both $\Delta T (M^{\rm b}_{\rm rem}) \leq E_{\rm EM}$ and $M^{\rm b}_{\rm rem} \leq M_{\rm th}$.

One example of this analysis is illustrated in Fig.~\ref{fig:example}.
Clearly, $E_{\rm EM}$ is so much smaller than $\Delta T_{\max}$ that the extractable energy curve intersects $E_{\rm EM}$ at the very precipice of the SMNS-HMNS transition.
We also find for all the EOS we have examined that $M^{\rm b}_{\rm smns} \approx 1.18M^{\rm b}_{\rm max}$ largely irrespective of compactness,
consistent with previous findings (e.g. \citealt{Lasota+96}).  
These two facts allow formulation of an approximate analytic criterion on the maximal non-rotating NS mass consistent with \LIGO,
\begin{equation} \label{eq:analytic_condition1}
M^{\rm b}_{\rm max} \lesssim M^{\rm b}_{\rm rem} / \xi ,
\end{equation}
where $\xi \simeq 1.16-1.21$ and the EOS is only necessary in translating baryonic to gravitational masses.

In addition to several key properties of each EOS, Table \ref{table:1} provides the probability that each EOS is consistent with constraints from \LIGO.  For instance, the very hard MS1, MPA1 and ENG EOS are disfavored, with consistencies of 0.0\%, 0.0\% and 5.2\%, respectively.  However, the softer EOS's with $M_{\rm max}^{\rm g} \lesssim 2.1-2.2M_{\odot}$ show much higher consistencies.  Figure \ref{fig:eos_survey} shows where each of our EOS lie in this $M_{\rm max}-R_{1.3}$ parameter space, with the strength of the symbol representing the probability of its consistency with \LIGO.  
Additionally shown are consistency values for polytropic EOSs of the form $p \propto \rho^{1+1/n}$ with indices $n= 0.5, 0.6, 0.7$. 
These define diagonal curves in the $M^{\rm g}_{\rm max}$---$R_{1.3}$ plane
parameterized by the pressure normalization of the polytrope.
Regions of large compactness are ruled-out by the requirement of causality, $R_{\rm max} \geq 2.82 G M^{\rm g}_{\rm max} / c^2$ (dark shaded region; \citealt{Koranda+97}) which is conservative since $R_{1.3}$ is generally larger than the radius of a maximum mass NS, $R_{\rm max}$. A tighter estimate of $R_{1.3} \geq 3.1 G M^{\rm g}_{\rm max} / c^2$ is therefore also shown (light shaded region).
The background grey curve shows the cumulative probability distribution function that the maximum mass $M_{\rm max}$ is less than a given value.  This was calculated by marginalizing over the $R_{1.3}$ axis and treating the consistency values as points in a probability distribution function.
We weight EOS with $M^{\rm g}_{\rm max}$ below $2.01M_{\odot}$ by a Gaussian prior accounting for
consistency with the maximum measured pulsar mass of $2.01\pm 0.04M_{\odot}$ \citep{Antoniadis+13}.
Thus, we find $M^{\rm g}_{\rm max} \lesssim 2.17M_{\odot}$ at 90\% confidence.

\section{Discussion}
\label{sec:discussion}

Several works have explored the potentially exotic EM signals of BNS mergers in cases when a long-lived SMNS or stable neutron star remnant is formed  \citep{Metzger+08,Bucciantini+12,Yu+13,Metzger&Bower14,Metzger&Piro14,Gao+16,Siegel&Ciolfi16a,Siegel&Ciolfi16b}.  However, one of the biggest lessons from \LIGO~was the {\it well-behaved} nature of its EM emission, under the simplest case of a relatively short-lived HMNS remnant \citep{Shibata&Taniguchi06}, with an off-axis afterglow \citep{vanEerten&MacFadyen11} and KN emission \citep{Metzger+10} closely resembling ``vanilla" theoretical predictions.

Here we have made explicit the argument that the BNS merger \LIGO~formed a HMNS.  In combination with the GW-measured binary mass, this inferred outcome places upper and lower limits on the maximum NS mass.  The lower limit on $M_{\rm max}$ is not constraining compared to those from well-measured pulsar masses, though tighter constraints would be possible by the future detection of similar fast-expanding blue KN ejecta (indicating HMNS formation) from a future BNS merger with a similar observing inclination but higher binary mass than \LIGO.  The lack of a luminous blue KN following the short GRB050509b \citep{Bloom+06,Metzger+10,Fong+17} may implicate a high mass binary and prompt collapse for this event.\footnote{A prompt collapse might also be consistent with the low measured gamma-ray fluence of GRB050509b, because the mass of the remnant accretion torus responsible for powering the GRB jet would also be lower for a prompt collapse than if a HMNS had formed.}

On the other hand, our upper limits on $M_{\rm max} \lesssim  2.17 M_{\odot}$ (90\% confidence limit; Fig.~\ref{fig:eos_survey}) are more constraining than the previous weak upper limits from causality, and less model-dependent than other methods (e.g.~\citealt{Lasky+14,Lawrence+15,Fryer+15,Gao+16,Alsing+17}).  
A low value of $M_{\rm max}$ has also been suggested
based on Galactic NS radius measurements
\citep{Ozel+16,Ozel&Freire16} and would be consistent with the relatively small NS radius $\lesssim 11$ km inferred from modeling the blue KN \citep{Nicholl+17,Cowperthwaite+17}.  
Furthermore, the lack of measurable tidal-effects in the inspiral of \LIGO~similarly imply a small NS radius and thus a low $M_{\rm max}$ \citep{LIGO+17DISCOVERY}.
Upper limits on $M_{\rm max}$ will be improved by the future discovery of EM emission from a merger with a lower total mass than \LIGO.  Conversely, the detection of a substantially brighter afterglow or faster evolving KN emission could instead point to the formation of a long-lived SMNS or stable remnant.   The NS masses measured for \LIGO~are broadly consistent with being drawn from Galactic NS population, which is well-fit by a Gaussian of mean $\mu = 1.32M_{\odot}$ and standard deviation $\sigma = 0.11M_{\odot}$ \citep{Kiziltan+13}; this hints that the HMNS formation inferred in \LIGO~is likely a common$-$if not the most frequent$-$outcome of a BNS merger. 

A simple analytic estimate of our result can be obtained from 
Eq.~(\ref{eq:analytic_condition1}),
using the approximation
$M_{\rm b} = M_{\rm g} + 0.075 M_{\rm g}^2$
for the relation between baryonic and gravitational masses \citep{Timmes+96}.
From this relation, the total baryonic binary mass is constrained, $M^{\rm b}_{\rm rem} \lesssim M^{\rm b}_{\rm tot} \lesssim 3.06M_\odot$.
A typical value of
$\xi \approx 1.18$ then implies that
\begin{equation}
M^{\rm g}_{\rm max} \lesssim \frac{ \sqrt[]{1 + 0.3 M^{\rm b}_{\rm rem} / \xi} - 1 }{0.15} \lesssim 2.2 M_\odot ,
\end{equation}
in agreement with our more elaborately calculated result.
We stress that 
the calculation above is intended only as an approximate analytic estimate, 
and that we do {\it not} use the \cite{Timmes+96} relation nor do we assume a universal value for $\xi$ in our complete analysis (\S~\ref{sec:EOS}).

Our approach differs in several respects from similar works constraining $M_{\rm max}$ \citep[e.g.][]{Lawrence+15,Fryer+15}.
These works generally assume 
(a) that creation of a GRB implies a BH formed, and (b) that BH formation necessarily implies either prompt-collapse or a HMNS post-merger remnant.
The central engine and emission mechanisms of GRBs are still widely debated in the literature, and the validity of the ``GRB=BH'' assumption remains unclear. 
Baryon-pollution by neutrino-driven winds launched off a long-lived NS remnant may hinder ultra-relativistic jets \citep{Murguia-Berthier+14}, however the Lorentz factors of short GRB jets
and GRB 170817A in particular
are poorly constrained, 
and there remains room for the possibility that short GRBs may be powered by strongly-magnetized rapidly-rotating NSs. 
NS GRB engines have also been suggested on grounds of the `extended' X-ray emission observed after some short GRBs \citep[e.g.][]{Metzger+08,Rowlinson+13}, emission which is difficult to interpret within the BH engine model.
Furthermore, the peculiar properties of GRB 170817A accompanying \LIGO,
although broadly consistent with a normal GRB viewed off-axis \citep[e.g.][]{Margutti+17}, 
may also point at a difference between this event and cosmological short GRBs \citep[e.g.][]{Gottlieb+17}, necessitating further caution in the GRB modeling and interpretation.
Secondly, even if a BH did form as a prerequisite to the GRB in this event (i.e. within $\sim2$s post-merger), there is nothing a-priori preventing this from occurring through the spin-down induced collapse of a SMNS merger remnant, negating assumption (b) above.
Here we have circumvented both assumptions 
and GRB engine modeling
altogether by instead relying on a simple energetic consideration --- that a SMNS merger remnant would inevitably release an enormous amount of rotational energy into the surrounding KN ejecta and circum-stellar medium.
Therefore, only merger remnants with an extractable rotational energy $\lesssim 10^{51} \, {\rm erg}$ are consistent with the energetics inferred from EM observations of \LIGO.

Several uncertainties affect our conclusions.  Our upper limits on $M_{\rm max}$ implicitly assume that this is the most important parameter controlling the HMNS-SMNS boundary, and that the suite of EOS we have taken are sufficiently ``representative" in the requisite sense.  We cannot obviously exclude the possibility that an alternative EOS could be found with large $M_{\rm max}$ that would still be consistent with our observational constraints.  

Another uncertainty affecting our conclusions is the possibility that in counting the KN and GRB components of the ejecta, we are somehow ``missing" substantial additional energy imparted by a putative SMNS remnant to the environment; however, any such hidden ejecta should be at least mildly relativistic and thus tightly constrained by radio synchrotron emission on timescales of months to years following the merger \citep{Metzger&Bower14}.  Yet another uncertainty is the possibility that a SMNS did form, but most of its rotational energy was lost to GW radiation instead of being transferred to the merger ejecta.  Though unlikely, we found this would only be possible for remnant lifetimes of $\sim 100 \,{\rm s}$ (Fig.~\ref{fig:Bfields}).
Searches in the \LIGO~waveform have revealed no evidence for such signals, although the detectors' decreasing sensitivity at high-frequencies currently limits these constraints \citep{LVC17PostMerger}.

Finally, our analysis neglects the effects of thermal pressure on the stability of the SMNS \citep{Kaplan+14}, which can be important on timescales of hundreds of milliseconds to seconds post merger (depending also on the effects of neutrino-driven convection; \citealt{Roberts+12}). Thermal pressure in the outer layers of the star generally acts to reduce the maximum mass of the SMNS remnant by up to $\lesssim 8$\% (mainly by reducing the angular velocity at the mass-shedding limit), which would act to weaken our constraints on $M_{\rm max}$.  Future numerical work exploring the transition from the HMNS to SMNS phase, which includes the effects of neutrino cooling and convection self-consistently, is required to better understand how this would quantitatively affect our conclusions.

\acknowledgements
BM and BDM are supported in part by NASA through the ATP program, grant numbers NNX16AB30G and NNX17AK43G.
The statistical analysis in this work would have greatly benefited from open access to the best-fit posterior parameter distributions obtained by \cite{LIGO+17DISCOVERY}.

\bibliographystyle{yahapj}

\end{document}